\documentclass[reprint,
superscriptaddress,
amsmath,amssymb,
aps,prd,floatfix,
twocolumn,
showpacs,
]{revtex4-2}

\usepackage{amsthm}
\usepackage{tikz}
\usepackage{graphicx}
\usepackage{dcolumn}
\usepackage{bm}
\usepackage{braket}
\usepackage[english]{babel}
\usepackage{hyperref}
\usepackage{amsmath}
\usepackage{graphicx}
\usepackage{float}
\usepackage{siunitx}
\usepackage{xcolor}
\usepackage{xspace}
\usepackage{eucal}
\usepackage{ulem}
\usepackage{multirow}
\usepackage[utf8]{inputenc}
\usepackage{pgfplots}
\pgfplotsset{compat=1.14}
\usepackage{upgreek}
\usepackage{siunitx}
\usepackage{gensymb}
\usepackage{amsmath}
\usepackage[justification=justified,singlelinecheck=false]{caption}
\usepackage{subcaption}
\usepackage{textcomp}
\usepackage{comment}
\usepackage{cleveref}

\Crefname{figure}{Figure}{Figures}

\usepackage{todonotes}
\usepackage{tabularx}
\usepackage{booktabs}
\usepackage{comment}
\DeclareSIUnit\g{g}
\DeclareSIUnit\gal{Gal}
\DeclareSIUnit\torr{Torr}
\DeclareSIUnit\bar{Bar}
\DeclareSIUnit\kelvin{K}
\DeclareSIUnit\inch{inch}
\DeclareSIUnit\joule{J}
\DeclareSIUnit\rad{rad}

\renewcommand{\t}[1]{\mathrm{#1}}

\begin{document}

\title{Optomechanical cooling and inertial sensing at low frequencies}

\author{Yanqi Zhang}
\affiliation{Texas A\&M University, Aerospace Engineering \& Physics, College Station, TX 77843}
\affiliation{James C. Wyant College of Optical Sciences, University of Arizona, 
\\1630 E. University Blvd., Tucson, AZ 85721}
\author{Adam Hines}
\affiliation{Texas A\&M University, Aerospace Engineering \& Physics, College Station, TX 77843}
\author{Dalziel Wilson}
\affiliation{James C. Wyant College of Optical Sciences, University of Arizona, 
\\1630 E. University Blvd., Tucson, AZ 85721}

\author{Felipe Guzman}
 \email{felipe@tamu.edu}
\affiliation{Texas A\&M University, Aerospace Engineering \& Physics, College Station, TX 77843}
\date{\today}

\begin{abstract}
An inertial sensor design is proposed in this paper to achieve high sensitivity and large dynamic range in the sub-Hz frequency regime. High acceleration sensitivity is obtained by combining optical cavity readout systems with monolithically fabricated mechanical resonators. A high-sensitivity heterodyne interferometer simultaneously monitors the test mass with an extensive dynamic range for low-stiffness resonators.  The bandwidth is tuned by optical feedback cooling to the test mass via radiation pressure interaction using an intensity-modulated laser. The transfer gain of the feedback system is analyzed to optimize system parameters towards the minimum cooling temperature that can be achieved. To practically implement the inertial sensor, we propose a cascaded cooling mechanism to improve cooling efficiency while operating at low optical power levels. The overall system layout presents an integrated design that is compact and lightweight.
\end{abstract}

\maketitle

\section{Introduction}

Acceleration sensing is crucial for tasks such as seismology\,\cite{seismology-app} and inertial navigation\,\cite{inertial-navi}, as well as a myriad of applications spanning the automobile\,\cite{automobile}, aerospace\,\cite{aerospace-app}, and consumer electronics industry\,\cite{acsaelm2021}. In the past decade, advances in cavity optomechanics\,\cite{cavity-optomech} have made it possible to develop accelerometers with ultrahigh sensitivity\,\cite{Krause2012,Guzman2014,ligo,aligo2015,darkmatter2021}, relevant to tasks such as geodesy\,\cite{Hines2022,Abich:2019cci}, gravitational wave detection\,\cite{ligo,aligo2015} and dark matter searches\,\cite{darkmatter2021}. Such optomechanical accelerometers typically consist of a mechanical resonator whose test mass oscillates in response to external accelerations. They also include an optical cavity to enhance the radiation pressure interaction between electromagnetic fields of light and modes of the mechanical resonator. This interaction provides high sensitivity measurements of the test mass displacement\,\cite{cavity-enhanced-book}.
It also provides the ability to tune the resonator frequency and damping rate via dynamic radiation pressure back-action\,\cite{optical-spring-1,optical-spring-2}, enabling control over the accelerometer bandwidth and dynamic range.

In the low-frequency regime (below \SI{1}{Hz}), the development of high sensitivity optomechanical accelerometers faces several challenges. The first challenge is to combine high displacement sensitivity with a large dynamic range; for resonators with low resonant frequency, the test mass displacement can reach amplitudes of hundreds to thousands of microns in 
applications such as seismometry. In this case, cavity-enhanced readout systems such as Fabry-Perot interferometers (FPIs) are not suitable, since their dynamic range is typically lower than the optical wavelength. 
In addition, various noise sources become significant in the low-frequency regime, such as laser frequency and thermo-elastic noise\,\cite{Nofrarias:2013uwa,Gibert:2014wga}. Alternative readout systems with high dynamic range have been explored, such as the heterodyne laser interferometer proposed for the Laser Interferometry Space Antenna (LISA) and the interferometer launched in LISA Pathfinder,\cite{Heinzel:2004sr, Aston:2006AIPC, Schuldt:2006SPIE, Guzman:2009thesis}. The LISA Pathfinder interferometer has achieved a displacement sensitivity of $\SI{10}{pm/\sqrt{Hz}}$ at $\SI{1}{mHz}$ on ground, and $\SI{30}{fm/\sqrt{Hz}}$ in space.
However, its assembly involves complicated alignment and bonding techniques. Recently we have developed compact common-mode heterodyne interferometers\,\cite{Joo:2020JOSAA,Zhang:2021,Zhang:22,Zhang:22_OE} for inertial sensing that provide high sensitivity. However, the noise floor of the interferometer is a few orders of magnitude higher than the thermal noise of the test mass, pointing toward the need for a hybrid approach.

A second challenge is that lowering the resonant frequency of the mechanical resonator entails adding mass or reducing stiffness, resulting in a bulky and likely delicate system. Moreover, radiation pressure back-action damping becomes less efficient for low-frequency resonators, as it entails operating in the ``bad cavity limit"\,\cite{BRAGINSKY2001, Arcizet2006, Gigan2006, Schliesser2008}, where the mechanical frequency is much smaller than the cavity bandwidth. Active radiation pressure feedback damping, where the test mass motion is suppressed by derivative feedback onto the laser intensity, has been demonstrated as an effective method for gram-scale resonators with mechanical frequencies on the order of 100\,Hz or above\,\cite{Poggio2007, Corbitt2007, Corbitt2007b, MowLowry2008}. However, for large-mass resonators, this entails large radiation pressure forces and concomitantly high optical power handling capacity.

In this paper, we propose an optomechanical accelerometer capable of high performance at sub-Hz frequencies. A key feature is the integration of two readout systems: an FPI designed for high-sensitivity displacement measurements and a heterodyne interferometer designed for high dynamic range displacement measurements. We show how combining these approaches with radiation pressure enables cascaded feedback cooling, a strategy whereby the resonator's motion can be suppressed to within the linewidth of the FPI, allowing high sensitivity, high dynamic range, and reduced optical power requirements.
The paper is organized as follows: in Section~\ref{sec:2} we present the overall system design, including detailed analyses of the mechanical resonator, the optical readout systems, and the feedback cooling system; in Section~\ref{sec:3} we describe the feedback cooling strategy and how to optimize the feedback gain to maximize cooling efficiency; Section~\ref{sec:4} focuses on the practical implementation of the sensor, designing the system parameters based on an optimized feedback gain. Finally, we present our cascaded feedback cooling strategy and show that it relaxes the laser power requirements while maintaining the lowest effective temperature that can be reached by the cooling process. 

\section{System design}
\label{sec:2}
\subsection{Overall system layout}

\begin{figure}[htbp]
\includegraphics[width=.8\columnwidth]{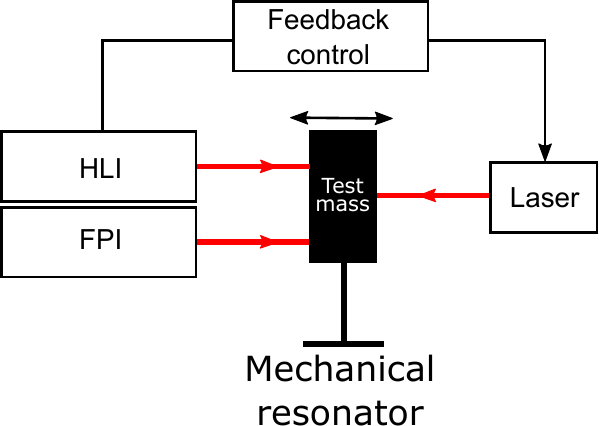}
\caption{\label{fig:overall-sketch}Sketch of the low-frequency inertial sensor design. Two optical readout systems measure the test mass displacement at the same time. The Fabry-Perot interferometer (FPI) performs displacement measurements with high sensitivity and serves as the main optical readout system. If the test mass motion is beyond the measurement range of FPI, the resonator oscillation is cooled down by the feedback control system via radiation pressure interaction. The heterodyne laser interferometer (HLI) obtains the test mass motion and feeds into the feedback control system.}
\end{figure}

The design concept for our low-frequency optomechanical inertial sensor is shown in Figure~\ref{fig:overall-sketch}. 
In our system, the acceleration test mass is suspended from a rigid frame (representing the inertial reference frame), forming a mechanical resonator.  The acceleration, $a$, is obtained by measuring the displacement $x$ of the test mass. The transfer function relating these two quantities is
\begin{equation}\label{eq:1}
    \frac{x(\omega)}{a(\omega)}=\frac{-1}{\omega_0^2-\omega^2+i (\omega\gamma_\text{v}+\omega_0^2\phi(\omega))},
\end{equation}
where $\omega_0$ is the resonance frequency of the test mass, $\gamma_\text{v}$ is its velocity damping rate (e.g., due to gas damping), and $\omega_0^2\phi(\omega_0)/\omega$ is its damping rate due to internal loss in the suspension, characterized by the loss coefficient $\phi(\omega)$\,\cite{Hines:2020qdi}.

Two optical readout systems monitor the test mass displacement simultaneously. An FPI with enhanced sensitivity serves as the main readout system, and its output is converted to acceleration using Equation~\ref{eq:1}.

To overcome the limited dynamic range of the FPI (smaller than the optical wavelength, $\lambda$), the test mass is simultaneously monitored over a significantly larger range by a heterodyne laser interferometer (HLI). The purpose of the HLI is to determine if the test mass displacement is within the FPI dynamic range. If not, then the HLI output is imprinted onto the intensity of an auxiliary laser field to implement radiation pressure feedback cooling.

\subsection{Mechanical resonator design and characterization}\label{sec:system-resonator}
\begin{figure}[htbp]
\includegraphics[width=0.8\columnwidth]{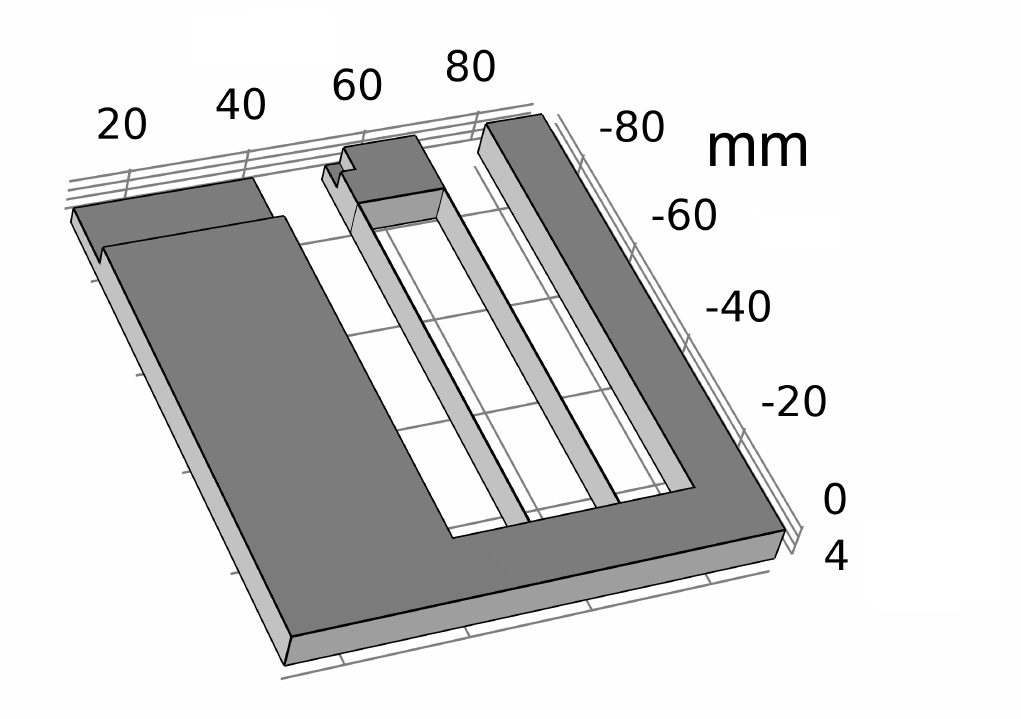}
\caption{\label{fig:reso-geo}Geometry of the mechanical resonator design. The \SI{2.6}{\gram} test mass is supported by two \SI{100}{\um}-thick flexures. The overall dimension is \SI{80}{\mm} $\times$ \SI{90}{mm} $\times$ \SI{6.6}{mm} and the total mass is \SI{58.2}{\g}. The notch on the baseplate is designed to integrate the optical readout system, and the notch on the test mass is to mount a plane mirror.}
\end{figure}

The mechanical resonator, depicted in Figure~\ref{fig:reso-geo}, consists of a \SI{2.6}{\gram} test mass supported by a pair of flexures with \SI{100}{\um} thickness. The overall footprint is \SI{80}{\mm} $\times$ \SI{90}{\mm} and the total mass is \SI{58.2}{\g}. The design, optimization, and performance analysis of a mechanical resonator with a similar geometry are discussed in detail in\,\cite{Hines:2020qdi}. The monolithic resonator is fabricated of a single fused silica wafer to minimize internal losses at room temperature\,\cite{FSloss}. 

The thermal motion of the test mass fundamentally limits the achievable acceleration sensitivity to 
\begin{align}
\label{eq:2}
    a_\mathrm{th}(\omega)&=\sqrt{\frac{4k_\mathrm{B}T}{m}\left(\gamma_\text{v}+\frac{\omega_0^2\phi(\omega)}{\omega}\right)}\\
    &= \sqrt{\frac{4k_\mathrm{B}T\omega_0}{m}\left(\frac{1}{Q_\text{v}}+\frac{1}{Q_\text{int}}\frac{\omega_0}{\omega}\frac{\phi(\omega)}{\phi(\omega_0)}\right)},
\end{align}

where $k_\mathrm{B}$ is the Boltzmann constant, $T$ is the device temperature,  $Q_\text{v}=\omega_0/\gamma_\text{v}$ is the mechanical quality factor ($Q$) due to gas damping and $Q_\text{int}= 1/\phi(\omega_0)$ is the mechanical $Q$ due to internal damping.

\begin{figure}[htbp]
\centering\includegraphics[width=0.5\textwidth]{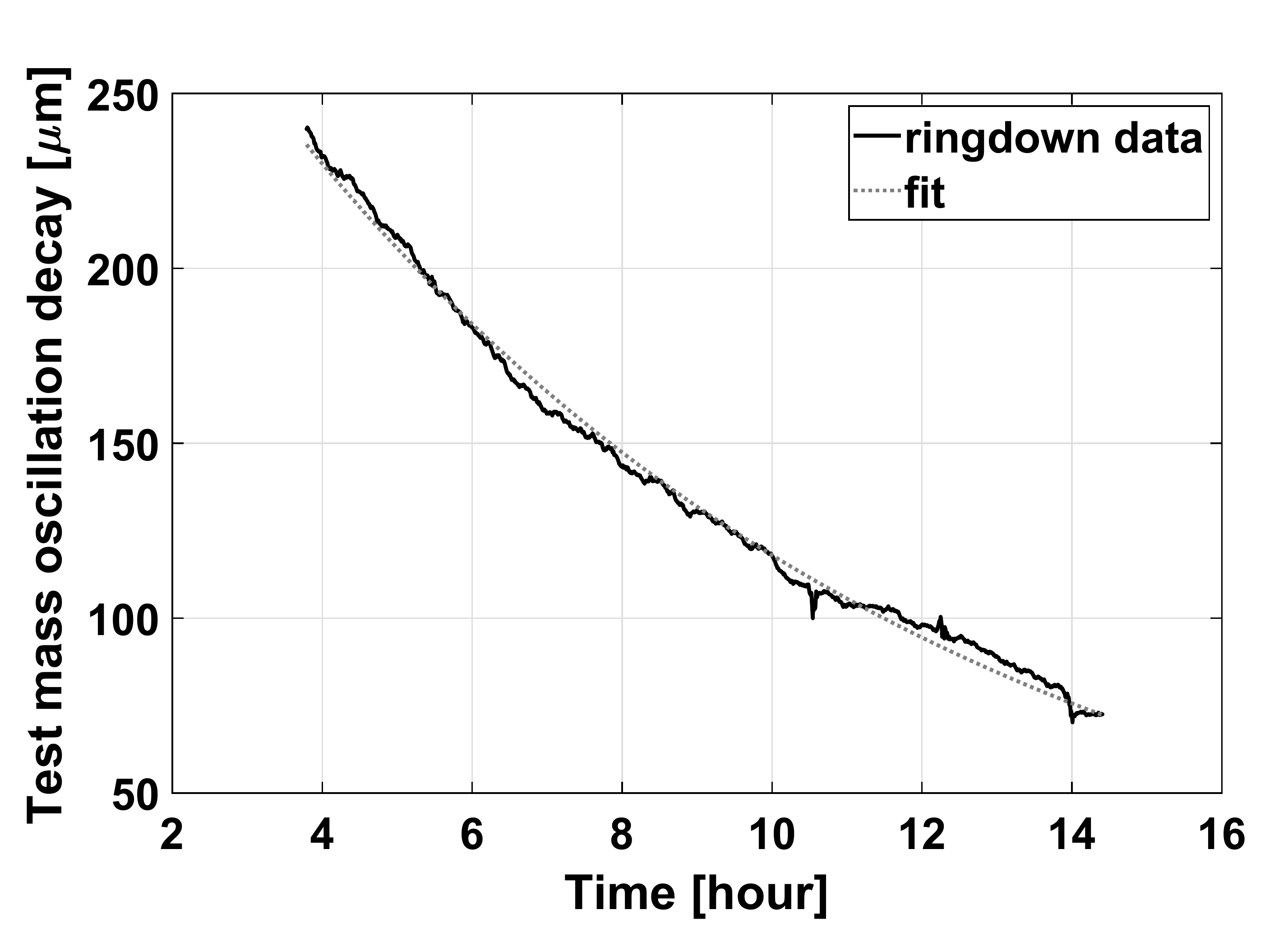}
\caption{\label{fig:ringdown}The envelope of the test mass ringdown measurement under vacuum conditions of $\SI{10}{\mu\torr}$. By fitting the envelope with an exponential decay, we obtain the quality factor $Q=\SI{4.77e5}{}$, corresponding to a $mQ$ product of \SI{1240}{kg}\,\cite{Hines2022}.}
\end{figure}

Equation \ref{eq:2} shows that a high value for $mQ/\omega_0$ is desirable to achieve a low thermal noise floor. To determine the resonance frequency $\omega_0$ and estimate the internal $Q$, we conducted ringdown measurements at a vacuum pressure of $\SI{10}{\upmu\torr}$ (to minimize gas damping, $\gamma_\text{v}$). The ringdown shown in Figure~\ref{fig:ringdown} is for a device with $\omega_0 = 2\pi\times\SI{4.72}{Hz}$, yielding a $Q=\SI{4.77e5}{}$, a $mQ$ product of \SI{1240}{kg}, and an acceleration noise floor of $a_\mathrm{th}(\omega_0)=\SI{1e-11}{m\,s^{-2}/\sqrt{Hz}}$ near resonance\,\cite{Hines2022}.

\subsection{Optical readout systems}

\subsubsection{Fabry-Perot Interferometer (FPI)}
For our main optical readout system, we envision a FPI of $L\approx 50$\,mm, formed by mounting one mirror onto the test mass and one onto the resonator frame. A shorter fiber-cavity approach has been previously demonstrated with stiffer resonator designs\,\cite{Guzman2014}; here the large dynamic range of our sensor requires a free space approach. The FPI is probed with a $\lambda = \SI{1064}{nm}$ laser whose frequency is locked to an FPI resonance using the Pound-Drever-Hall (PDH) technique \cite{Drever:1983qsr}. Meanwhile, the absolute laser frequency is tracked, for instance, by beating it against an iodine-referenced laser or another suitable optical frequency reference. Within the bandwidth of frequency measurement instrument, the cavity length $L$ and laser frequency $\nu$ fluctuations are related by
\begin{equation}\label{eq:4}
\frac{\nu}{L}=\frac{\bar{\nu}}{\bar{L}}=\frac{c}{\overline{\lambda}\overline{L}}
\end{equation}

Hence, the readout signal can be expressed as
\begin{subequations}\begin{align}
    \nu(\omega) &=  \frac{c}{\overline{\lambda}\overline{L}}\frac{a_\text{ext}(\omega)+ a_\text{th}(\omega)}{\omega_0^2-\omega^2+i (\omega\gamma_\text{v}+\omega_0^2\phi(\omega))} + \nu_\text{n}(\omega)\\
    &=  \frac{c}{\overline{\lambda}\overline{L}}\frac{a_\text{ext}(\omega)+ a_\text{th}(\omega)+ a_\text{n}(\omega)}{\omega_0^2-\omega^2+i (\omega\gamma_\text{v}+\omega_0^2\phi(\omega))},\label{eq:5}
\end{align}\end{subequations}

where $a_\text{ext}$ is the physical acceleration of the resonator (the desired signal), $a_\text{th}$ is the apparent acceleration of the resonator due to the thermal motion of the test mass (Eq. \ref{eq:2}), and $a_\text{n}$ is the apparent acceleration due to readout noise $\nu_\text{n}$. The latter arises from various sources including laser frequency noise, shot noise\,\cite{Black:2001B}, technical noise, and cavity mirror noise (e.g. thermo-elastic noise).

For slow changes, the dynamic range $\Delta L$, which is the largest measurable length change of the optical readout system, is determined by the laser frequency tuning range $\Delta \nu$:
\begin{subequations}\label{eq:7}\begin{align}
\Delta L &= \frac{\overline{\lambda}\overline{L}}{c} \Delta \nu\\
&= \SI{1.8}{\um}\times \frac{\Delta \nu}{\SI{10}{GHz}}\frac{\overline{\lambda}}{\SI{1064}{nm}}\frac{\overline{L}}{\SI{50}{mm}}
\end{align}\end{subequations}
In Eq. \ref{eq:7}b we estimate a dynamic range of 1.8\,$\upmu$m with a typical commercial Nd:YAG laser with a tuning range of \SI{10}{GHz}. We anticipate that this dynamic range will be too small for operation in a non-isolated environment since an equivalent test mass displacement would be produced by an equivalent acceleration input of only $\Delta L\omega_0/\sqrt{Q}\approx 8\;\text{n}g/\sqrt{\text{Hz}}$. With this in mind, in the following sections, we propose a HLI to monitor the full range of the test mass motion and a radiation-pressure feedback control system to damp the test mass using the HLI output as an error signal. 

\subsubsection{Long-range heterodyne laser interferometer (HLI)}
\begin{figure}[t]
\centering\includegraphics[width=0.9\columnwidth]{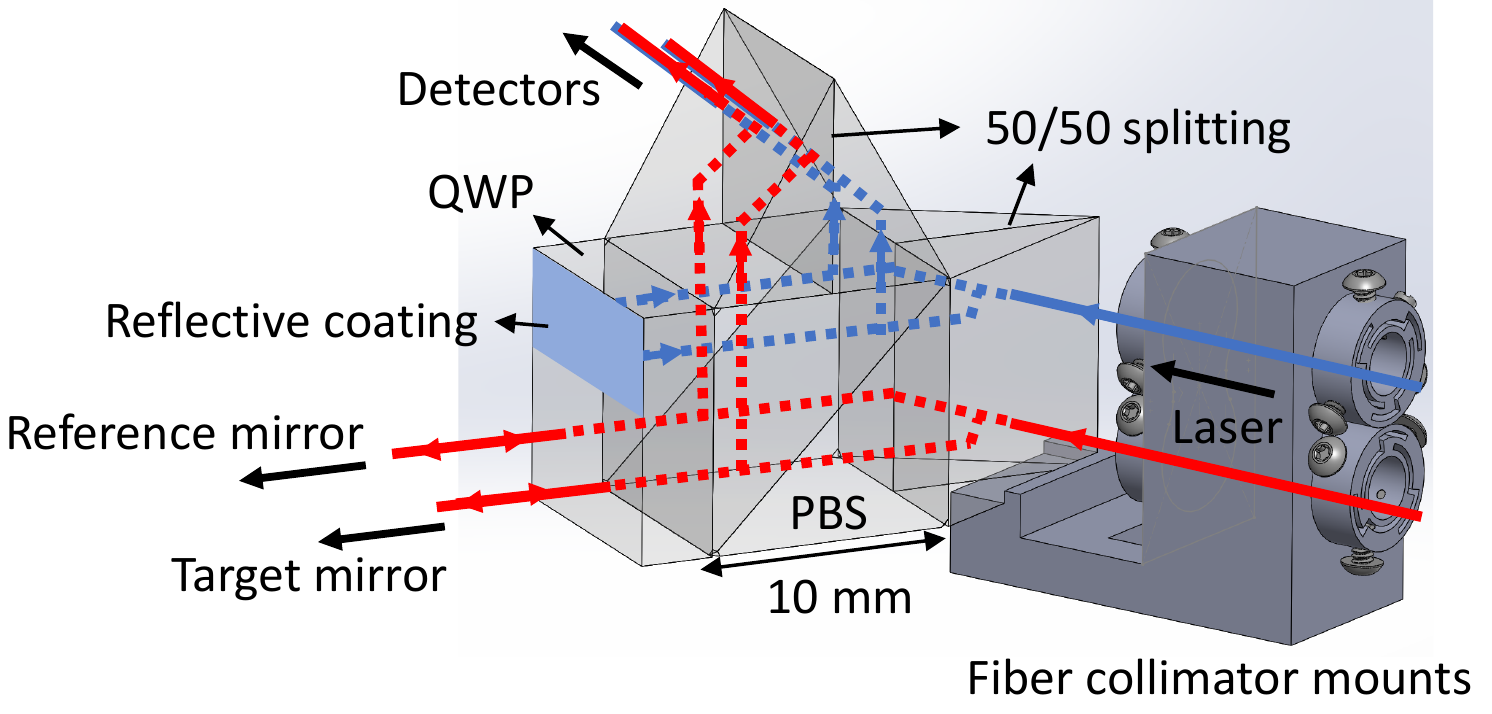}
\caption{Schematic diagram of the quasi-monolithic interferometer unit in the isometric view. Two incoming laser beams are split into four beams by the 50-50 non-polarizing splitting surface embedded in the equilateral triangular prism. The beam pair with different frequencies constructs one interferometer. The measurement interferometer (MIFO) measures the target displacement. The reference interferometer (RIFO) measures the systematic noises that share the common optical paths with the MIFO. The target displacement is calculated from the differential phase measurement between MIFO and RIFO. \label{fig:hli}}
\end{figure}

Heterodyne interferometry is a common technique that has the potential to achieve high sensitivity in addition to a large dynamic range and traceable calibration. Recently we have developed a compact HLI\,\cite{Zhang:22} for use with low-frequency optomechanical inertial sensors. 
Our HLI employs a common-path design, as shown in Figure~\ref{fig:hli}, to provide a high rejection ratio to various forms of environmental noise.

\begin{figure}[t]
\centering\includegraphics[width=0.48\textwidth]{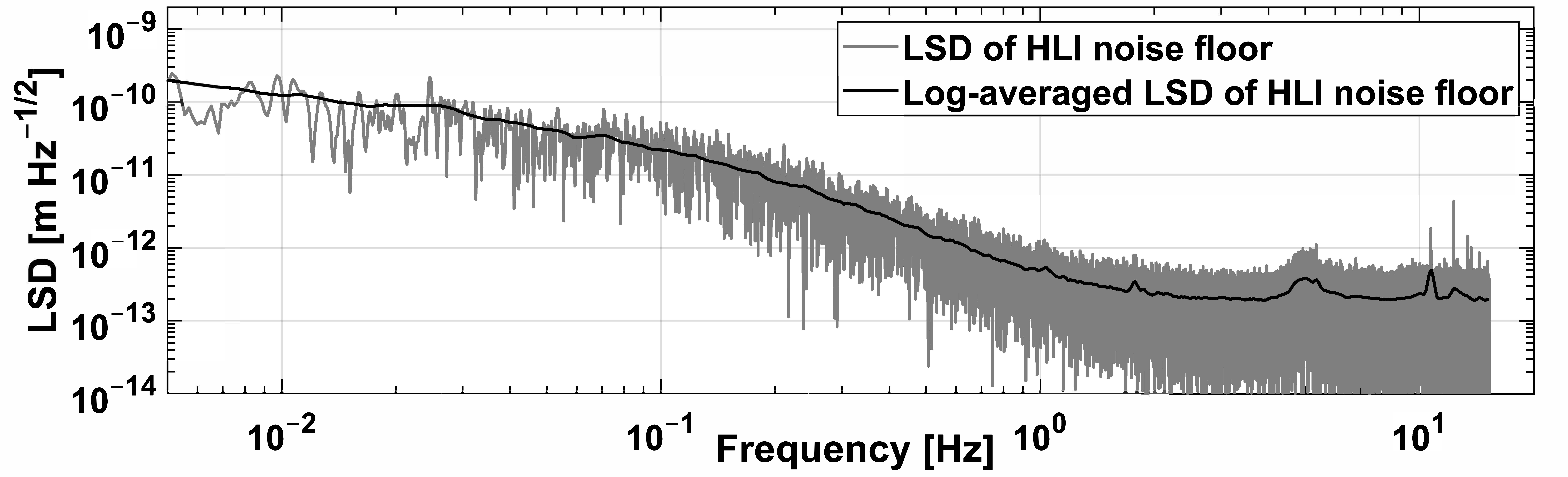}
\caption{\label{fig:lr-dmi}The linear spectral density (LSD) and its logarithmic average of the heterodyne laser interferometer (HLI)~\cite{Zhang:22} noise floor. The interferometer prototype shows a noise floor of \SI{2e-13}{m/\sqrt{Hz}} above \SI{1}{Hz} when tested in vacuum.}
\end{figure}

Concretely, a measurement interferometer (MIFO) measures the test mass motion, and a reference interferometer (RIFO) of common optical paths with MIFO, monitors ambient noise. The RIFO signal is then subtracted from the MIFO output to reduce its noise content. The complete system has a footprint of $\SI{20}{mm}\times\SI{20}{mm}\times\SI{10}{mm}$ and weighs \SI{4.5}{g}. To achieve such a compact assembly, all the optical components are cemented as a quasi-monolithic unit. The small size and weight enable integration onto the mechanical resonator's frame. Preliminary tests have yielded a noise floor of \SI{2e-13}{m/\sqrt{Hz}} around \SI{1}{Hz}, limited by photodetector noise, as shown in Figure~\ref{fig:lr-dmi}\,\cite{Zhang:22}. If the displacement recorded by the HLI is outside the dynamic range of the FPI, the HLI data is used as the error signal of the feedback control system described in the next section. 

\subsection{Feedback control system}\label{sec:system-feedback}

The feedback control system includes signal processing modules and digital-analog conversion interfaces to actuate the test mass. In our case, minimal contact with the test mass is desirable to reduce surface losses.  We therefore consider radiation pressure as a feedback actuator.  A further advantage of this approach is the traceability of the radiation pressure force through the laser wavelength\,\cite{Wagner:18,Melcher:14}.

\begin{figure}[h]
\centering
\includegraphics[width=0.9\columnwidth]{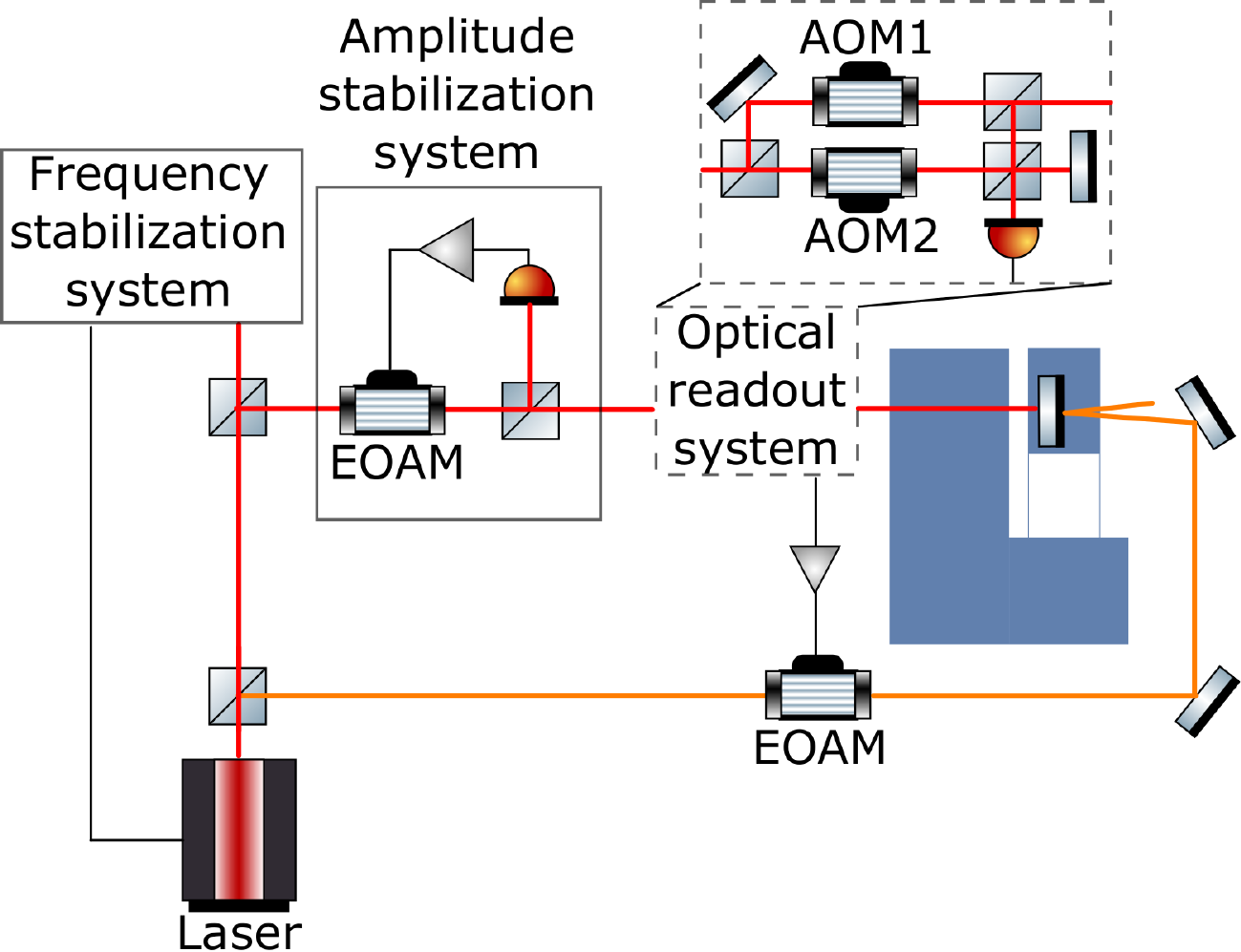}
\caption{\label{fig:fb-system}Layout of the feedback control system. The feedback force is provided by radiation pressure interaction between an intensity-modulated laser beam and the test mass. The laser intensity is modulated by an electro-optical amplitude modulator (EOAM). The displacement of the test mass is read out by a heterodyne interferometer (simplified in the figure), and sent to the feedback control loop. The controller in the feedback loop calculates and outputs a corresponding voltage signal to the EOAM to modulate actuating laser intensity. The laser is stabilized in frequency and amplitude to reduce the measurement noise.}
\end{figure}

Figure~\ref{fig:fb-system} shows our proposed radiation pressure feedback scheme, where we launch an auxiliary laser beam onto the test mass, whose intensity is modulated with the output of an HLI displacement sensor. To control the test mass position $x$ in a traceable manner, it is necessary to understand the transduction chain comprising the feedback circuit.  To this end we consider the following model
\begin{align}\label{eq:feedbackmodel}
x(\omega) &= \chi_\t{m}(\omega)(F_\t{th}(\omega)+F_\t{ext}(\omega)+F_\t{fb}(\omega))\\ \nonumber
&=\chi_\t{m}(\omega)(F_\t{th}(\omega)+F_\t{ext}(\omega)-\chi_\t{fb}(\omega)y(\omega))\\ \nonumber
&=\chi_\t{eff}(\omega)(F_\t{th}(\omega)+F_\t{ext}(\omega)-\chi_\t{fb}(\omega)x_n(\omega))
\end{align}
where $y = x+x_n$ is the apparent position detected by the HLI, with a readout noise $x_n$, $F_\t{fb}$ is the radiation pressure feedback force, $\chi_\t{fb}(\omega) \equiv - F_\t{fb}(\omega)/y(\omega)$ is the feedback gain, and 
\begin{equation}
\chi_\t{eff}(\omega) \equiv \frac{\chi_\t{m}(\omega)}{1+\chi_\t{m}(\omega)\chi_\t{fb}(\omega)}
\label{eqn:8}
\end{equation}
is the effective (closed-loop) mechanical susceptibility.

It is evident from Equation~\ref{eqn:8} that the thermal force sensitivity remains unchanged by the feedback actuation; however, as shown below, feedback damping ($\chi_\t{fb}\propto i x$) is advantageous because it allows the sensor to operate in the linear regime $x\ll\lambda/\mathcal{F}$, by reducing the displacement on resonance. The cost of this extended dynamic range is additional stochastic force due to the feedback of measurement noise, $\chi_\t{fb(\omega)}\,x_n(\omega)$.

The phase and magnitude of the feedback gain must be tailored to achieve a desired closed-loop response and noise figure, which can be tailored in the digital loop software. We thus consider the following model for the feedback gain
\begin{equation}\label{eq:12}
\chi_\t{fb}(\omega)=G_{FP}\,G_{PV}\,G_{Vx}(\omega),
\end{equation}
where $G_{FP}$, and $G_{PV}$ are the steady-state response functions of the radiation pressure actuator and intensity modulator, respectively, and $G_{Vx}(\omega)$ is the frequency dependent response function of the software-interfaced HLI that considers the digital processing for phase extraction (phasemeter).

To model the actuator transfer function, we assume the test mass has perfect reflectivity for the power $P$ incident on the test mass. Thus, we obtain
\begin{equation}
G_{FP}=\frac{dF_\t{RP}}{dP} = \frac{2}{c}.
\end{equation}

The transfer function of the intensity modulator depends on the modulation method.  We envision an intensity modulator based on a polarization-based EOAM, for which the output power $P$ (referred to the power incident on the test mass) is related to the voltage $V$ applied across the electrodes as $P = P_0 \cos^2(\pi V/V_\pi)$, where $P_0$ is the maximum transmitted power and $V_\pi$ is the half-wave voltage of the EOAM.  This leads to a transfer function
\begin{equation}
G_{PV}=\frac{dP}{dV} = \frac{\pi P_0 \sin(V/V_\pi)}{V_\pi}.
\end{equation}

Finally, the transfer function of the HLI depends on the phase measurement technique. Common phase extraction algorithms include phase-locked loop (PLL)~\cite{pll-book} and single-bin DFT algorithms~\cite{Heinzel:2004sr}. A disadvantage of these methods is that both involve computationally intensive inverse trigonometric functions, which lead to a non-linear time-invariant (NLTI) system. Instead, we propose a digital phasemeter, which only requires a low-pass filter. The phase response of the entire system is determined by this filter, and has a negligible effect at low frequencies. With this assumption, the HLI transfer function $G_{Vx}$ can be expressed as 
\begin{equation}\label{eq:16}
    G_{Vx}(\omega) = \frac{V(\omega)}{x(\omega)} = G_\t{DAC}(\omega)\frac{2\pi}{\lambda},
\end{equation}
where $G_\t{DAC}$ is an amplification factor that can be set when converting the measured phase to the analog voltage output applied to the EOAM. The significance of $G_\t{DAC}$ will be explained in detail in Section~\ref{sec:4-1}. 

Combining Equations~\ref{eq:12}-\ref{eq:16}, the feedback gain of the system shown in Figure~\ref{fig:fb-system}, can be expressed as
\begin{equation}\label{eq:17}
    \chi_\t{fb}(\omega) \approx \frac{4\pi^2}{c\lambda} \frac{G_\t{DAC}(\omega)P_0\sin{2\theta}}{V_\pi},
\end{equation}
where system parameters such as the laser power $P_0$ can be optimized to enhance the feedback cooling efficiency. 

\section{Feedback cooling optimization}
\label{sec:3}
\subsection{Optical cooling}
\label{sec:3-1}
In feedback cooling protocols, derivative feedback gain --i.e. a velocity-proportional feedback force-- is used to damp  the test mass displacement $x$ to the measurement noise floor $x_n$.  In practice, however, the feedback is bandwidth limited for the low frequencies considered here, such that digital filtering can provide a good approximation to a purely derivative gain.  We thus consider the following model
\begin{equation}\label{eq:18}
    \chi_\mathrm{fb}(\omega) = i\,m\,g\,\gamma_\t{m}(\omega)\omega,
\end{equation}
where $g$ is a unitless gain factor and $\gamma_\t{m}(\omega) = \gamma_\text{v}+\omega_0^2\phi(\omega)/\omega$ is a frequency depending damping term including the effects of viscous (gas) damping and internal damping, as in Equation~\ref{eq:1}.

Hence, the closed-loop susceptibility of the system is
\begin{subequations}\label{eq:19}\begin{align}
   \chi_\mathrm{eff}(\omega) &= \frac{1}{m[\omega_0^2-\omega^2+i(1+g)\gamma_\mathrm{m}(\omega)\omega]}\\&   \equiv \frac{1}{m[\omega_0^2-\omega^2+i\gamma_\t{eff}(\omega)\omega]},
\end{align}\end{subequations}
which is characterized by an effective damping rate $\gamma_\t{eff} \equiv (1+g)\gamma_\t{m}(\omega)$.

\begin{figure}[t!]
\includegraphics[width=0.9\columnwidth]{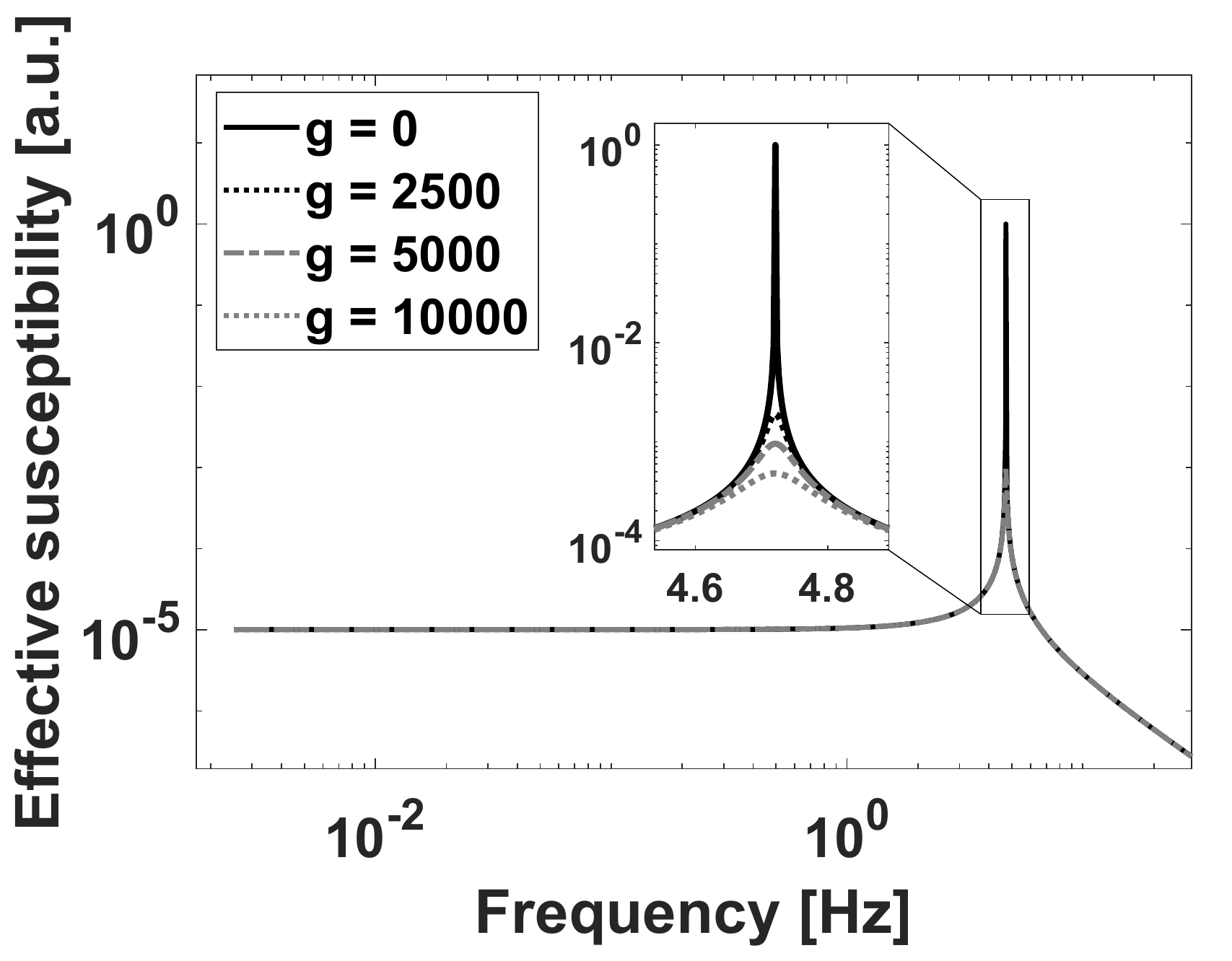}
\caption{\label{fig:eff-sus}Effective susceptibility $\chi_\mathrm{eff}(\omega)$ of the test mass for different gain factors $g=$ 0, 2500, 5000, and 10000. The case of $g=0$ represents the open loop mechanical susceptibility. Increasing $g$ reduces driven displacement on resonance. }
\end{figure}

\subsection{Feedback gain optimization}
Figure~\ref{fig:eff-sus} suggests that increasing the feedback gain $g$ reduces the test mass displacement at resonance. However, it also increases the feedback of the measurement noise, expressed by the term $\chi_\t{fb}x_n$ in Equation \ref{eq:feedbackmodel}.  As a result, there is an optimal gain at which the total closed-loop displacement can be minimized.  This can be seen by re-writing Equation \ref{eq:feedbackmodel} in terms of the closed loop displacement power spectral density
\begin{eqnarray}\label{eq:20}
    S_{xx}(\omega) &= |\chi_\mathrm{eff}(\omega)|^2\large(S_{FF}^\mathrm{th}(\omega)+S_{FF}^\mathrm{ext}(\omega)\\ \nonumber
    &+\left|\chi_\mathrm{fb}(\omega) \right|^2 S_{xx}^\mathrm{n}(\omega)\large)
\end{eqnarray}
and integrating over the resonance peak to give the closed loop displacement variance
\begin{equation}\label{eq:varianceclosedloop}
    \langle x^2\rangle\approx \frac{1}{1+g}\langle x_\t{th,0}^2\rangle+\frac{g^2}{1+g}\langle x_\t{n}^2\rangle + \langle x_\t{ext}\rangle^2(g)
\end{equation}
where $\langle x_\t{th,0}^2\rangle = k_B T/m\omega_0^2 \approx \gamma_\t{m} S_{xx}^\t{th}(\omega_0)/4$ is the open loop thermal displacement, $\langle x_\t{n}^2\rangle \approx \gamma_\t{m} S_{xx}^\t{n}(\omega_0)/4$ is the apparent displacement due to the measurement noise, and $\langle x_\t{ext}^2\rangle = \int |\chi_\t{eff}(\omega)|^2 S_\t{FF}^\t{ext}(\omega)d\omega/2\pi$ is the closed loop displacement due to external forces.

In principle, $g$ can be tailored to minimize $\langle x^2\rangle$. However, the magnitude of this gain depends on the external force spectrum $S_{FF}^\t{ext}(\omega)$.  Without knowing $S_{FF}^\t{ext}(\omega)$, it is customary to optimze $g$ to minimize the test mass displacement at resonance, i.e.
\begin{equation}\label{eq:22}
    g_\mathrm{opt} \approx \sqrt{\frac{\langle x_\t{th,0}^2\rangle}{\langle x_\t{n}^2\rangle}}= \sqrt{\frac{4k_\mathrm{B}T}{m\omega_0^2 \Gamma_\mathrm{m} S_{xx}^\mathrm{n}}},
\end{equation}
yielding 
\begin{equation}\label{eq:23}
    \langle x^2\rangle_\t{min} \ge 2\sqrt{\langle x_\t{th,0}^2\rangle\langle x_\t{n}^2\rangle}.
\end{equation}
Here we assume that $g_\t{opt}\gg 1$ corresponds to a large thermal-to-readout-noise ratio.  

Equivalently, the effect of feedback damping is to reduce the effective temperature of the test mass to
\begin{equation}
    T_\t{eff} = \frac{m\omega_0^2\langle x^2\rangle}{k_B} \ge \frac{1}{1+g}T+\frac{g^2}{1+g}T_n \ge 2\sqrt{T\,T_n},
\end{equation}
 where $T_n \equiv m\omega_0^2\langle x_n^2\rangle/k_B$ is the apparent temperature of the readout noise and the minimal value is obtained for $g=g_\t{opt}$ in the absence of external forces.

\begin{figure}
    \centering
    \subfloat[]{\includegraphics[width = 0.48\textwidth]{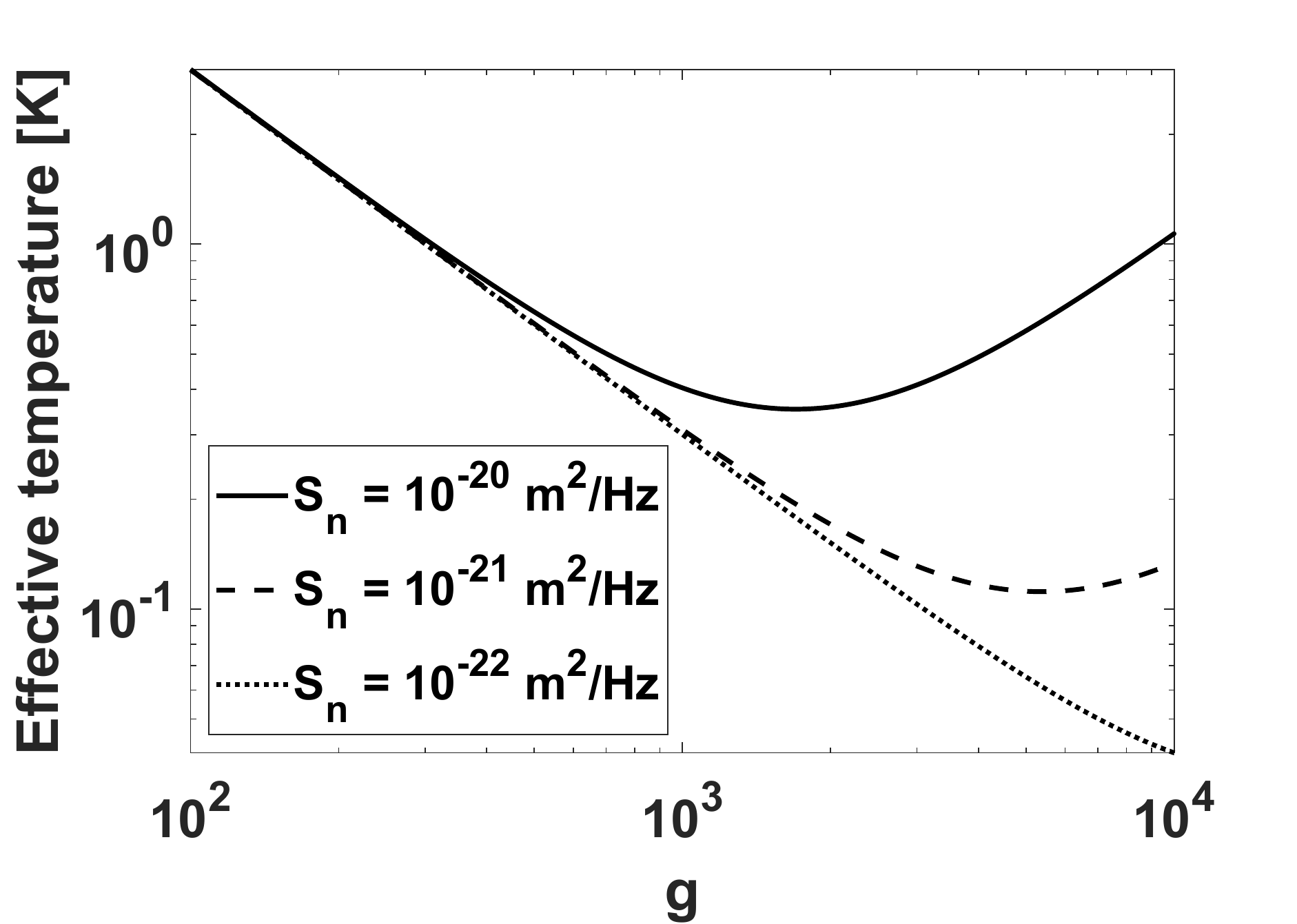}\label{fig:Teff}} \\
    \subfloat[]{\includegraphics[width = 0.48\textwidth]{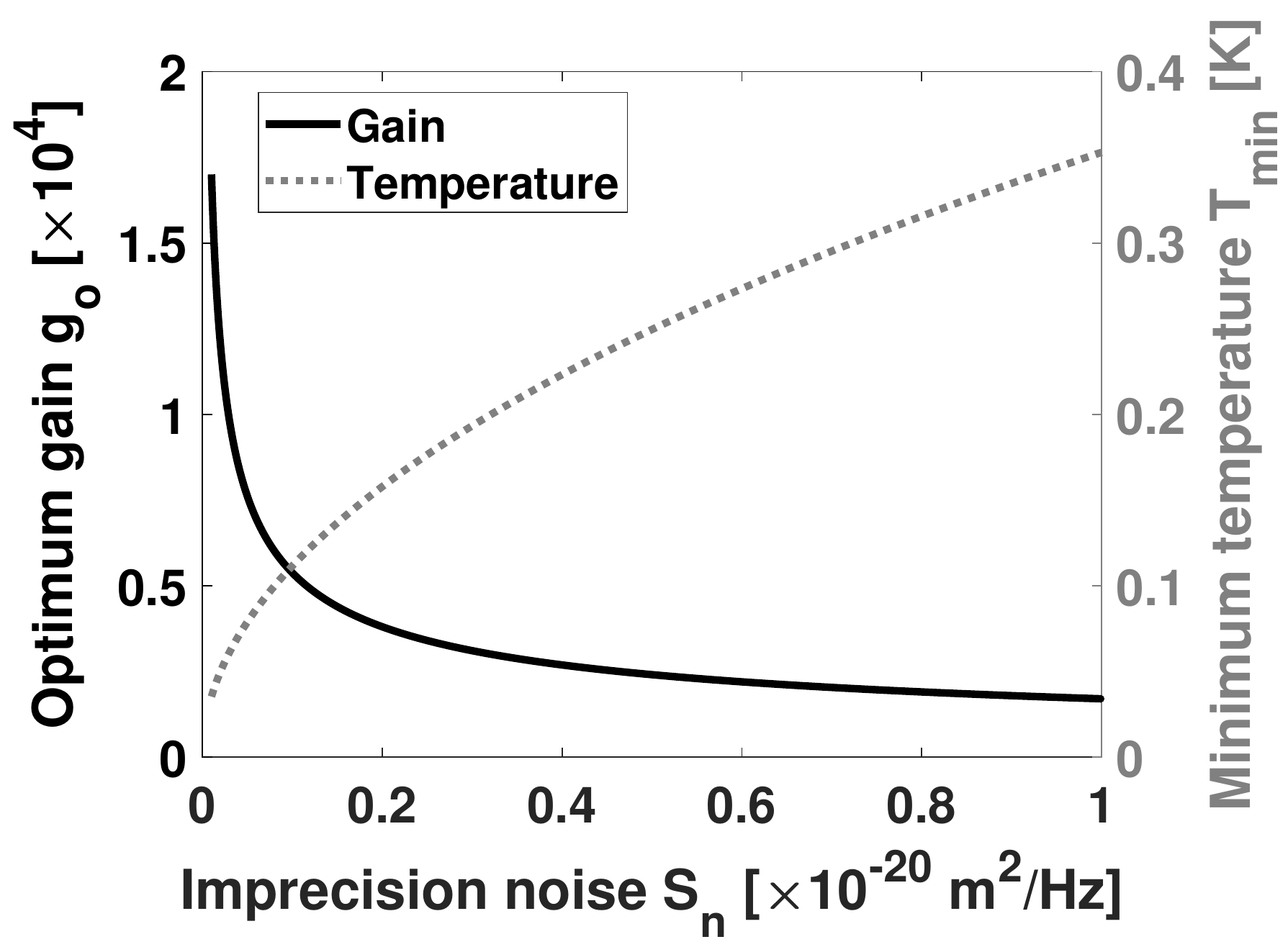}\label{fig:go}}
    \caption{(a) Relation between effective temperature and gain factor for different levels of readout imprecision noise; (b) The minimum temperature that the cooling system can achieve and the corresponding gain factor to achieve this temperature versus the readout imprecision noise in the system.}
    \label{fig:my_label}
\end{figure}

In Figure~\ref{fig:Teff}, we plot the effective temperature of the resonator described in Section~\ref{sec:system-resonator}, subject to feedback cooling with a variety of gains and measurement noise levels, assuming $T = 300$ K. In Figure~\ref{fig:go} we plot the optimal gain and associated minimal temperature as a function of measurement noise.  In the next section we consider a parameter space that is practically accessible.

\section{Feedback implementation}
\label{sec:4}
In this section, we consider how to practically implement feedback cooling of our low frequency resonator, described in Section~\ref{sec:system-resonator}\,\cite{Hines2022}.  We are particularly interested in the requirements to cool the test mass to the measurement noise floor, corresponding to a gain factor of $g = g_\t{opt}$ (Equation \ref{eq:22}).  Building upon the recent development of a monolithic HLI integrated with a similar test mass \cite{Zhang:2021}, we conservatively assume $\sqrt{S_{xx}^\mathrm{n}} = \SI{5e-12}{m/\sqrt{Hz}}$. Combined with the test mass properties described in Section~\ref{sec:system-resonator}, this implies $g_\mathrm{opt} = \SI{3.40e4}{}$. We find that achieving this gain factor is challenging with the long-range (HLI) detection scheme, due to its (deliberately) small transduction gain.  To overcome this challenge, we propose a cascaded feedback cooling protocol employing both long-range (HLI) and high sensitivity (FPI) readout systems in series.

\subsection{System parameters}
\label{sec:4-1}

Combining Equations~\ref{eq:17} and~\ref{eq:18}, near the resonance frequency, the feedback gain factor $g$ is related to the system parameters in the control loop by
\begin{equation}\label{eq:25}
    g = |\chi_\t{fb}(\omega_0)| \frac{Q}{m\omega_0^2}
    = \frac{4\pi^2}{c \lambda} \frac{Q}{m\omega_0^2} \frac{|G_\t{DAC}(\omega_0)|P_0 \sin{2\theta}}{V_\pi}.
\end{equation}
In practice, the EOAM phase is adjusted to maximize the modulation depth, corresponding to $\sin{2\theta}=1$. The remaining free parameters are the optical power $P_0$, the half-wave voltage $V_\pi$, and the digital conversion gain $G_\t{DAC}$, which is the parameter that provides the largest tuning range.

As mentioned in Section~\ref{sec:system-feedback}, the digital gain, $G_\t{DAC}$, is a variable factor in the digital-to-analog conversion (DAC) coefficient that translates the digital phasemeter output — a phase in radian — into a corresponding proportional analog voltage that is injected into the EOAM to control the optical power used for feedback cooling. In practice, the magnitude of $G_\t{DAC}$ is limited by the EOAM $V_\pi$ and the physical displacement range of the test mass $x_\mathrm{p-p}$, where 

\begin{equation}
\label{eq:G-DAC-max}
    G_\t{DAC}^{\mathrm{max}}=\frac{V_\pi\lambda}{2\pi x_\mathrm{p-p}}. 
\end{equation}

\begin{figure*}[t!]

\includegraphics[width=.95\textwidth]{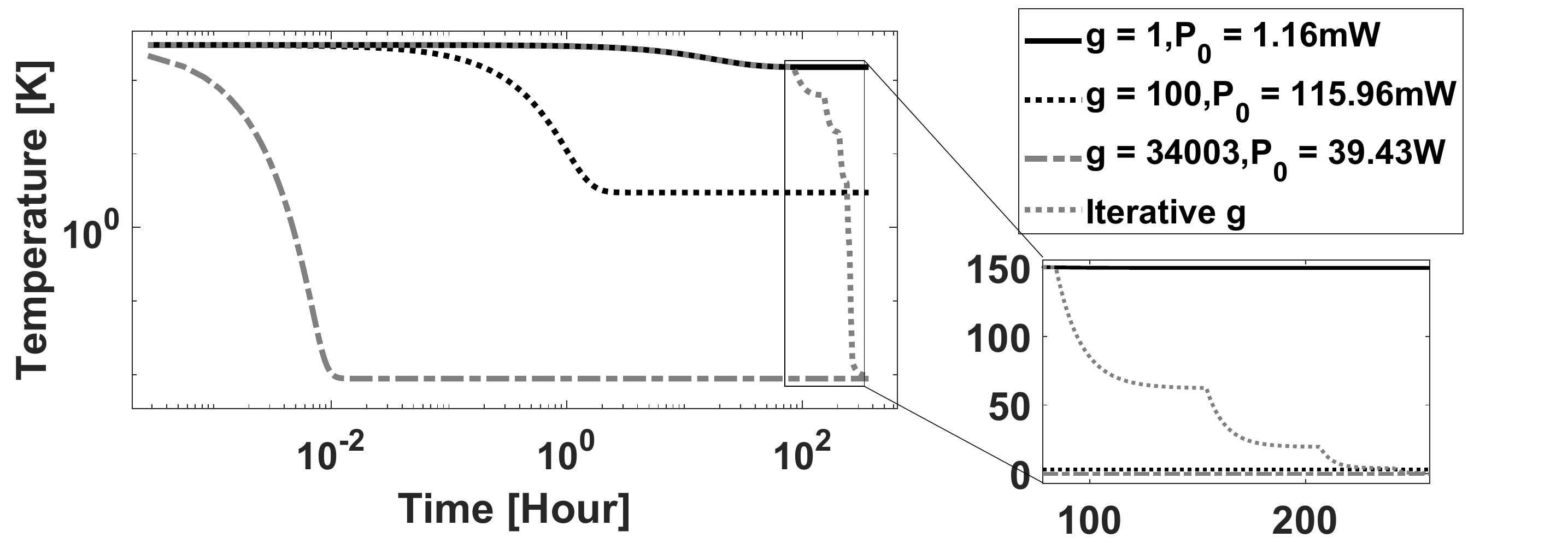}
\caption{\label{fig:cascade-cooling}Effective temperature of the test mass subject to feedback cooling with a constant gain factor $g$ = 1, 100, and $\SI{3.40e4}{}$, and for an iteratively increased $g$ (cascaded feedback cooling). Each step on the iterative curve corresponds to the moment when $G_\t{DAC}$ changes. The inset shows a zoom-in on the final state for $g$ = 100 and $\SI{3.40e4}{}$, and the iterative $g$. In the cascaded cooling mechanism, the initial gain factor is $g_0=1$. After 6 iterations, its final state reaches the same level as single-step feedback cooling with $g_\mathrm{opt} = \SI{3.40e4}{}$. For iterative cooling, the optical power is fixed at $P_0 = \SI{1.16}{mW}$. }
\end{figure*}

\subsection{Feedback cooling with HLI detector}

The open loop displacement of our test mass has been measured to be on the order of $\pm\SI{100}{\um}$ in our laboratory, without vibration isolation. Combining Equations \ref{eq:22}, \ref{eq:25}, and \ref{eq:G-DAC-max}, with the mechanical properties of the resonator in Section \ref{sec:system-resonator},  we predict that the minimum laser power $P_0$ needed to achieve a gain factor $g=1$ ($T_\t{eff} = T/2$) is $\SI{1.16}{mW}$, and the laser power to achieve $g_\mathrm{opt} = \SI{3.40e4}{}$ is $\SI{34.43}{W}$.  Thus we find that, for our current test mass, feedback cooling with the long range HLI detector is challenging due to the required high laser power.  In particular, $P_0\gtrsim 100$ mW has the potential to damage the EOAM crystal. It also poses a challenge in terms of thermal management and handling of other components in the system such as the fiber couplers and photodetectors. To reduce the power requirements, we propose a cascaded cooling mechanism employing a dynamical gain $G_\t{DAC}$, as discussed below.

\subsection{Cascaded feedback cooling}
The critical factor limiting the single stage cooling approach is the readout conversion coefficient $G_\t{DAC}$. We can use the fact that the maximum level of this gain, $G_\t{DAC}^\t{max}$, can go higher as the test mass is cooled, requiring less optical power. This means that we can \emph{dynamically} increase $G_\t{DAC}$ as the oscillation amplitude of the test mass goes down.  In fact, we can also dynamically switch from the low sensitivity HLI readout scheme to the high sensitivity FPI readout scheme when the test mass oscillation is suppressed to $\sqrt{\langle x^2\rangle}<\lambda/\mathcal{F}$, which also reduces backaction due to readout noise.  The implementation of this cascaded feedback cooling approach from the slow relaxation time of our $\gamma_\t{m}\sim 2\pi\cdot 10\,\mu\t{Hz}$ test mass, is given by~\cite{Pinard2000}, 
\begin{equation}\label{eq:26}
    \langle x^2(t) \rangle = \frac{\langle x^2(0)\rangle}{1+g}\left( 1+g\exp^{-(1+g)\gamma_\t{m}t}\right).
\end{equation}

\begin{figure}[htbp]
\centering
\includegraphics[width=\columnwidth]{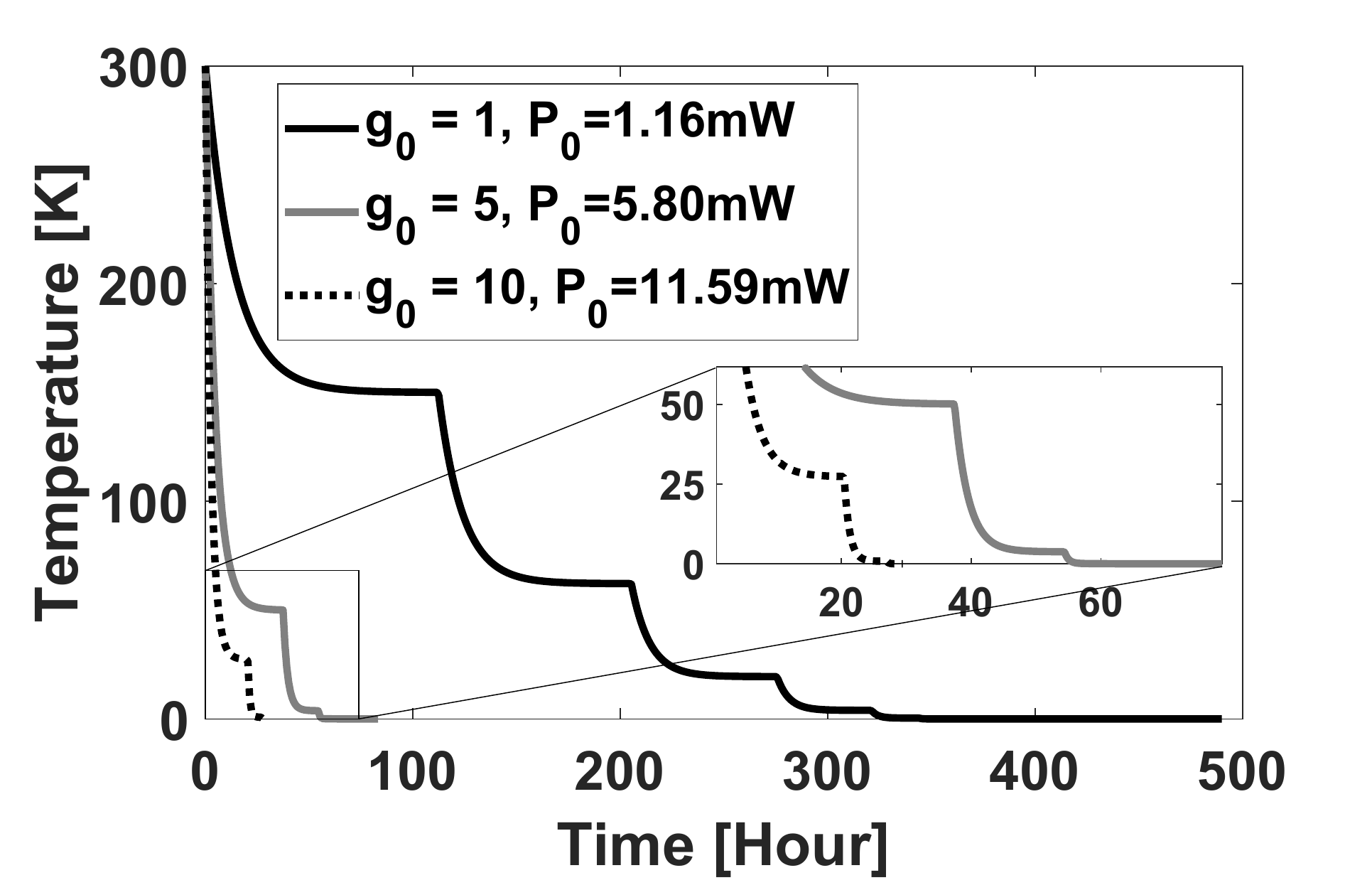}
\caption{\label{fig:diff-g0}Effective temperature in a cascaded cooling mechanism for different initial gain factors $g_0$ = 1, 5, and 10. A large $g_0$ allows the system to operate much faster and with less iterations to achieve a certain final state.}
\end{figure}

In Figure~\ref{fig:cascade-cooling} we present simulations of single-step and cascaded feedback cooling of our test mass for various gain settings and optical powers, using the HLI detector and system parameters described above. Relative to single-step cooling with $P_0 \approx 39,\t{W}$, cascaded cooling enables optimal gain ($g_\t{opt} = 3.4\cdot 10^4$), requiring $10^4$-fold lower optical power ($P_0\approx 1\,\t{mW}$), at the expense of a $10^4$-fold higher cooling time (10 days, versus 1 minute).  As shown in Figure~\ref{fig:diff-g0}, the total cooling time for the cascaded approach depends on the initial gain factor $g_0$, which is proportional to optical power, and scales approximately inversely with $g_0$.  In practice, however, we can cool the test mass down to within the FPI capture range at much lower gains than $g_\mathrm{opt}$, which allows us to switch to a FPI readout much faster. In this case, the required time in our cascaded cooling approach, can be reduced to a few hours using an optical power feedback of only 11.59\,mW, as shown in Figure~\ref{fig:diff-g0}.

\section{Complete system overview}
\begin{figure}[h]
\centering
\includegraphics[width=\columnwidth]{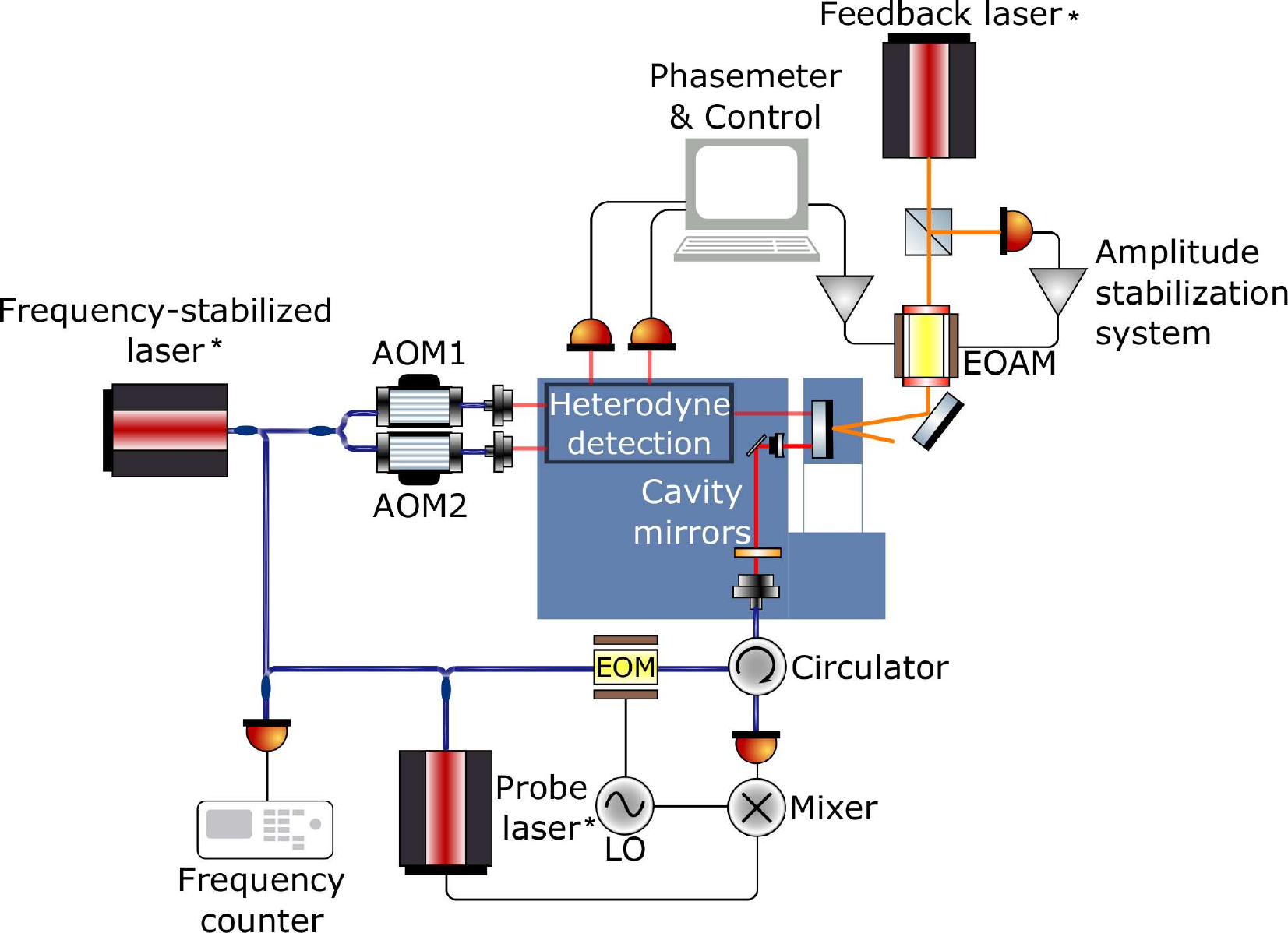}
\caption{\label{fig:overall}Overall layout of the optomechanical inertial sensor design, including details of the mechanical resonator, the optical readout systems, and the feedback control system. Three individual lasers are used in this figure due to their different purposes in the system.\\ 
\textbf{Note:} In a practical implementation, all three lasers can be replaced with a single laser source, following a proper optical setup and frequency management.}
\end{figure}

The complete system we propose is shown in Figure~\ref{fig:overall}, including the dual measurement scheme and radiation pressure feedback circuit. In this figure we show three lasers for simplicity, however, we note that all these three sources can be obtained from a single laser unit, following a proper optical setup and frequency management. The three optical sources are: a) the frequency-stabilized laser that serves as the optical source for the heterodyne interferometer to monitor the full-range motion of the test mass, as well as the frequency standard to obtain the frequency drift of the probe laser, b) the probe laser that is the optical source for the FPI as the main optical readout system to perform highly sensitive displacement measurements, and c) the feedback laser that is interfaced with an EOAM intensity modulator for feedback cooling.

The feedback circuit is shown at the top-right of Figure~\ref{fig:overall}. For simplicity, we omit the fact that the HLI output passes through a near-resonant bandpass filter and a phase shifter in this Figure, to avoid crosstalk from higher-order modes as well as to provide a $90\degree$ phase shift for the feedback force, respectively. The optical components of both the heterodyne interferometer and the FPI can be mounted on the resonator frame, leading to a compactly integrated and vacuum-compatible inertial sensor. This design also reduces the effect of mechanical vibrations and thermal expansion, which are significant noise sources at low frequencies. 

In closed loop, the detected laser frequency fluctuations in Equation~\ref{eq:5} can be rewritten as  
\begin{equation}\label{eq:27}
   \Delta \nu = \frac{c}{\overline{\lambda}\overline{L}}\frac{a_\text{ext}(\omega)+ a_\text{th}(\omega)+ a_\text{n}(\omega)}{\omega_0^2-\omega^2+i(1+g)\omega\omega_0 / Q}, 
\end{equation}
where $g$ in the cascaded cooling system can be traced to the final $G_\t{DAC}$ via Equation~\ref{eq:25}. All the parameters in Equation~\ref{eq:27}, such as $\omega_0$, $Q$, $\overline{\lambda}$, and $\overline{L}$ can be measured directly and independently from the displacement sensitivity measurement. 

\section{Conclusions and outlook}
\label{sec:5}
We have proposed an optomechanical inertial sensor that features high displacement sensitivity, to enable thermal-noise-limited performance, and the ability to perform large-amplitude acceleration measurements in non-isolated environments via radiation pressure feedback cooling. The design principles are described in detail for each sub-system, followed by a discussion of a practical implementation. A novel cascaded cooling approach is presented, which reduces the laser power requirements and improves the cooling efficiency.

Cascaded feedback cooling of our relatively large, low-frequency test mass allows the use of low optical power, at the milliwatt level. However, this feature comes at the expense of prolonging the time needed for the optical cooling process, when compared to a single-step method that utilizes high laser power (tens of Watts). Nonetheless, this extended time in reaching the minimal operational resonator temperature may likely be acceptable for applications where the sensor is deployed permanently, and the cascaded cooling is part of the initial calibration process. Seismic monitoring of a particular site or space missions are examples of such applications, since they typically require commissioning of their instrumentation over comparable time scales. Our design thus introduces a promising optomechanical inertial sensing technology that is capable of reaching high acceleration sensitivity at low frequencies (below $\SI{1e-9}{m\,s^{-2}/\sqrt{Hz}}$ above 1\,mHz\,\cite{Hines2022}). This system is currently under development in our laboratory. 

\subsection*{Funding}
The authors acknowledge financial support from the National Science Foundation (NSF) through grants PHY-2045579 and ECCS-1945832, and the National Aeronautics and Space Administration (NASA) through grant 80NSSC20K1723.

\bibliography{References}

\end{document}